\documentclass[12pt,preprint]{aastex}
\usepackage[numberedappendix]{emulateapj5}
\usepackage{amstext}
\usepackage{apjfonts}
\usepackage{amsmath}

\newcommand{\pii}{\left[{\cal P}_{2}\right]_{i}}

\newcommand{\bpi}{{\cal B}_i\left[{\cal P}_{2}\right]_{i}}
\newcommand{\cpi}{{\cal C}_i\left[{\cal P}_{2}\right]_{i}}
\newcommand{\dpi}{{\cal D}_i\left[{\cal P}_{2}\right]_{i}}
\newcommand{\epi}{{\cal E}_i\left[{\cal P}_{2}\right]_{i}}
\newcommand{\fpi}{{\cal F}_i\left[{\cal P}_{2}\right]_{i}}

\journalinfo {\underline{\bf To Appear in the Astrophysical Journal (\apj).}}
\shorttitle{BLACK HOLE ACCRETION WITH ISOTHERMAL SHOCK}
\shortauthors{{DAS}, {PENDHARKAR} \& {MITRA}}

\begin{document}


\title{Multi-transonic Black Hole Accretion Discs with Isothermal Standing
Shocks}


\author{Tapas K. \ Das $^{1,2}$, Jayant K. \ Pendharkar$^{3,4}$, Sanjit \ Mitra$^{3}$}
\affil{$^1$Division of Astronomy, Department of Physics and Astronomy,
University of California at Los
Angeles, Box 951562, Los Angeles, CA 90095-1562, USA\\
$^2$ Institute of Geophysics and Planetary Physics, University of California at Los
Angeles, Box 951567, Los Angeles, CA 90095-1567, USA\\
$^3$ IUCAA, Post Bag 4 Ganeshkhind,
Pune 410 007, India\\
$^4$ Department of Astronomy, Osmania University, Hyderabad 500 007, India.\\
e-mail: tapas@astro.ucla.edu, jkp@iucaa.ernet.in, sanjit@iucaa.ernet.in}
\begin{abstract}
\noindent
In this work we would like to address the issue of shock formation in
black hole accretion discs. We provide  a {\it generalized two parameter 
solution
scheme} for multi-transonic, accretion and wind around
Schwarzschild black holes, by mainly concentrating on accretion solutions
which may contain steady, standing isothermal shocks. 
Such shocks 
conserve flow temperature by the dissipation of energy at the shock
location. We use the vertically
integrated 1.5 dimensional model to describe the disc structure, where the
equations of motion apply to the equatorial plane of the central
accretor, assuming the flow to be in hydrostatic equilibrium in the transverse
direction. Unlike previous works in this field, our calculation is {\it not}
restricted to any particular kind of post-Newtonian gravitational potentials,
rather we use {\it all}
available pseudo-Schwarzschild potentials to formulate and
solve the equations governing the accretion and wind only in terms of 
flow temperature $T$ and specific angular momentum $\lambda$ of the flow.
The accretion flow
is assumed to be non-dissipative everywhere,
except possibly at the shock location, if
any. We observe that a significant region of parameter space spanned by 
$\left\{{\lambda},T\right\}$ allows shock formation. Our generalized
formalism assures that the shock formation is not just an artifact of a
particular type of gravitational potential, rather inclusion of all available
black-hole potentials demonstrates
a substantially extended zone of parameter space
allowing for the possibility of shock formation. We thus arrive at the
conclusion that the standing shocks are essential ingredients in
rotating, advective accretion flows of isothermal fluid around a non-spinning
astrophysical black hole. 
We identify {\it all} possible shock solutions which may be
present in isothermal disc accretion and thoroughly study the dependence of
various shock parameters on fundamental dynamical variables governing the
accretion flow for {\it all} possible initial boundary conditions.
Types of shocks discussed in this paper may appear to be
`bright' because of the huge amount of energy dissipation at shock,
and the quick removal of such energy to maintain the isothermality may 
power the strong X-ray flares recently observed to be emerged from
our Galactic centre.
 The results
are
discussed in connection to other astrophysical phenomena of related
interest, such as 
the QPO behaviour of galactic black hole candidates.
\end{abstract}




\section{Introduction}
\noindent
Perturbation of various kinds may produce discontinuities in astrophysical fluid flow. 
Discontinuities in fluid flow are said to take place over one or more surfaces when 
any dynamical and / or thermodynamic quantity may change discontinuously as such
surfaces are crossed; the corresponding surfaces are called surfaces of discontinuity.
Certain boundary conditions are to be satisfied across such surfaces and according to
those conditions, surfaces of discontinuities are classified into various categories; 
the most important are being the shock waves or shocks, which,
when the flow is adiabatic, obey the following
conditions (Landau \& Lifsitz, 1959, LL hereafter):
$$
\left[{\rho}u\right]=0,~
\left[p+{\rho}u^2\right]=0,~
\left[\frac{u^2}{2}+w\right]=0
$$
where $\rho, u, p$ and $w$ are the fluid density, flow velocity,
fluid pressure and the enthalpy of the flow respectively, and for
any two arbitrary flow variables $\alpha$ and $\beta$,
$$
\left[\alpha+\beta\right]\equiv
\left\{\left(\alpha+\beta \right)_{-}-\left(\alpha+\beta \right)_{+}\right\}
$$
$$
\left[\alpha\beta\right]\equiv\left\{{\alpha_-}{\beta_-}-{\alpha_+}{\beta_+}\right\}.
$$
The subscripts
$\left(+\right)$ and $\left(-\right)$ refer to  the quantities measured just after and just
 before the flow encounters the surface(s) of discontinuity. Quite often, such shock waves are generated in
 various kinds of supersonic astrophysical flows having
intrinsic angular momentum, resulting
in a flow which
becomes subsonic. This is because of the fact that the repulsive centrifugal potential barrier
 experienced by such flows are sufficiently strong to brake the motion and a stationary solution
 could be introduced only through a shock. Rotating, transonic astrophysical fluid flows are thus believed
 to be `prone' to the shock formation phenomena.
It has been established in recent years that
in order to
satisfy the inner boundary conditions imposed by the
event horizon, accretion onto black holes should
exhibit transonic properties in general; 
which further
indicates that formation of shock waves
are  possible in astrophysical fluid flows onto
galactic and extra-galactic black holes. One also
expects that shock formation in black hole accretion discs
might be a general phenomenon because shock waves
in rotating astrophysical flows are convincingly
able to provide an important and efficient mechanism
for conversion of a significant amount of the
gravitational energy (available from deep potential
wells created by these massive compact accretors) into
radiation by randomizing the directed infall motion of
the accreting fluid. Hence shocks possibly play an
important role in governing the overall dynamical and
radiative processes taking place in astrophysical fluids and 
plasma accreting 
onto black holes.
Thus the study of steady, standing, stationary shock waves produced in black
hole accretion has acquired a very important status in recent
years.\\

The issue of the formation of steady, standing shock waves in black hole accretion discs is addressed in
two different ways in general. First, one can study the formation of Rankine-Hugoniot shock waves (RHSW)
in a 
polytropic flow. Radiative cooling in this type of shock is quite inefficient. No energy is
dissipated at the shock and the total specific energy of the accreting material is a shock-conserved
quantity. Entropy is generated at the shock and the post-shock flow possesses
 a higher entropy accretion rate
than it's pre-shock counterpart. The flow changes its temperature permanently at the shock. Higher
post-shock temperature puffs up the post-shock flow and a quasi-spherical,
 quasi-toroidal centrifugal
pressure supported boundary layer is formed in the inner region of the accretion disc. 
Fukue (1987)
was the pioneer to initiate the systematic study of such shock dynamics
using a rigorous equation of state valid for from the non-relativistic regime to 
the highly relativistic regime.
Then Chakrabarti and collaborators (Chakrabarti 1989, 1996 and references therein, Abramowicz \&
Chakrabarti
1990, Chakrabarti and Molteni 1993) used the vertically integrated flow 
model of Matsumoto et. al. (1984) to extended Fukue's work towards the 
global study of shock formation in post-Newtonian geometry.
However, such shock solutions were
obtained either on a case
by case basis, or, even when successful attempts were made  to provide 
a more complete analysis,
the boundary of the parameter space
responsible for shock formation was obtained only for global variation of 
the total specific energy ${\cal E}$ (or accretion rate ${\dot M}$)
and specific angular momentum $\lambda$ of the flow, and not for
variations of
the polytropic constant $\gamma$ of the
flow; rather, accretion was usually considered to be 
ultra-relativistic\footnote{By the term `ultra-relativistic' and `purely non-relativistic'
we mean a flow with
$\gamma=\frac{4}{3}$ and $\gamma=\frac{5}{3}$ respectively,
according to the terminology used in
Frank et. al. 1992.}, 
which may {\it not} always be a realistic assumption.
As $\gamma$ is expected to 
have great influence on the radiative properties of the flow in general,
ignoring the explicit 
dependence of shock solutions on $\gamma$ limits claims of generality. 
Also,
all such so-called global shock solutions have been discussed 
only in the context of one particular type of pseudo-Schwarzschild 
black hole potential, namely,
the Paczy\'nski \& Wiita (1980) potential 
($\Phi_1$ hereafter).
For accretion in Kerr geometry, various authors (Lu 1985,1986, Lu et al 1987,
Nakayama and Fukue 1989) made attempts to provide a complete
general relativistic description of shocked accretion flow by making 
a number of assumptions, some of which, however,
may not appear to be fully convincing (see a detail discussion in Das 2002)
and it is fair to say that 
no well accepted complete
general relativistic global shock solution exclusively obtained
for a hydrodynamic accretion disc
around a Schwarzschild BH has yet appeared in the literature.
Motivated by these limitations, Das (2002, D02 hereafter) 
provided a generalized formalism allowing one to
construct and solve the equations governing the hydrodynamic black hole accretion of a
polytropic fluid in all
 available pseudo-Schwarzschild potentials which may contain steady, standing RHSW.
In this work,
the explicit dependence of
 shock solutions on all important accretion parameters was
thoroughly studied. The generalized formalism
 developed by D02 has also been used to explain related interesting phenomena
 like the
analytical computation
 of QPO frequencies (Das 2003) 
and the generation of accretion powered-galactic jets (Das, Rao and Vadawale
 2003).\\
In another class of shock study, one concentrates on shock formation in isothermal black hole accretion
discs. The characteristic features of such shocks are quite different from the
non-dissipative shocks discussed
above. In isothermal shocks, the
accretion flow is allowed to dissipate a part of its
energy and entropy at
the shock surface to keep the post-shock temperature equal to its pre-shock value. 
This maintains the vertical
thickness of the flow exactly the
same just before and just after the shock is formed. Simultaneous jumps in
energy and entropy join the pre-shock supersonic flow to its post-shock
subsonic counterpart. Chakrabarti (1989a, C89 hereafter) studied the
properties of such shocks in the rotating isothermal accretion around
compact objects. However, his work
was restricted to only one post-Newtonian black hole potential, i.e., the
Paczy\'nski and Wiita (1980) potential 
$\left(\Phi_{1}\right)$ and the dependence of shock parameters on initial boundary 
conditions was not thoroughly studied.
Also, the degeneracy of the shock locations was not removed completely. Later on, Yang and Kafatos (1995, YK
hereafter) investigated the formation of isothermal shocks and studied the stability properties of such shocks 
for
general relativistic accretion onto Schwarzschild black holes. They also, quite successfully, computed the
jump conditions for relativistic shocks. YK analyzed the stability of shocks in  a more efficient way
compared to the method used by C89 and could remove the two-fold degeneracy of shock locations; in this way
they could properly identify the most stable
shock location. However, unlike C89, YK used a 
one-dimensional conical model for accretion discs,
 where the existence of vertical hydrostatic equilibrium was
not taken into account. Also,
 the explicit dependence of various shock-related quantities on accretion
parameters was not fully explored. Nevertheless, YK still deserves 
special attention because it is the
only work in the literature which attempts to study 
isothermal shock formation in a fully relativistic
frame work.\\

In this paper, we would like to address the issue of shock formation in isothermal black
hole accretion discs from a more general point of view. Instead of confining our
calculation to a
particular black hole potential, {\it all} available pseudo-Schwarzschild potentials will
be exhaustively used to investigate the possibility of shock formation in hydrodynamic,
inviscid, isothermal black hole accretion discs. 
Following Matsumoto et. al. (1984), we will
be using the vertically integrated 1.5-dimensional model to describe the 
structure of the accretion discs, which is, perhaps,  more
realistic compared to simplified one dimensional conical flow model used by YK. In this
1.5 dimensional model (which has extensive use in literature, see Chakrabarti 1996 and references therein) 
the equations of motion are
written on the equatorial plane of the central accretor whereas the flow remains in hydrostatic
equilibrium in the transverse direction. For all black hole potentials we first formulate and solve
the equations governing multi-transonic accretion and wind around the central accretor
in terms of {\it only} two parameters, namely, specific flow angular momentum $\lambda$ and flow
temperature $T$. Then we provide a self-consistent scheme to capture the stable shock location(s)
for multi-transonic accretion flow and finally we thoroughly study the dependence of various
shock related quantities on fundamental accretion parameters. In this way we provide a
generalized two parameter solution scheme for multi-transonic isothermal black hole
accretion disc for all available pseudo-Scharzschild black hole potentials which contains
dissipative shock waves. Our generalized formalism assures that our model is not just an
artifact of a particular type of potential only, and inclusion of every black hole potential allows
a substantially extended zone of parameter space allowing for the possibility of shock
formation.\\ 
Rigorous investigation of the complete general relativistic
multi-transonic black hole accretion disc and wind
is believed to be extremely complicated.
At the same time it is
understood that as relativistic effects play an important role in the
regions close to the accreting black hole (where most of the
gravitational potential energy is released), purely Newtonian gravitational
potential (in the form ${\Phi}_{N}=-\frac{GM_{BH}}{r}$, where $M_{BH}$ 
is the mass of the compact accretor)
cannot be a realistic choice to describe
transonic black hole accretion in general. To compromise between the ease of
handling of a
Newtonian description of gravity and the realistic situations
described by complicated general relativistic calculations, a series of
`modified' Newtonian potentials have been introduced
to describe the general relativistic effects that are
most important for accretion disc structure around Schwarzschild 
black holes.
Introduction of such potentials allows one to investigate the
complicated physical processes taking place in disc accretion in a
semi-Newtonian framework by avoiding pure general relativistic calculations
so that
most of the features of spacetime around a compact object are retained and
some crucial properties of the analogous relativistic
solutions of disc structure could be reproduced with high accuracy.
Hence, those potentials might be designated as `pseudo-
Schwarzschild' potentials.
Our calculations in this paper will be based on four such pseudo
Schwarzschild potentials.
Explicit forms of those four potentials are
the following (see Das \& Sarkar 2001, D02 and 
Artemova, Bj\"{o}rnsson \& Novikov 1996, ABN hereafter, and
references therein, for detail discussion about various properties
of these potentials):
$$
\Phi_{1}(r)=-\frac{1}{2(r-1)}~;~
\Phi_{2}(r)=-\frac{1}{2r}\left[1-\frac{3}{2r}+12{\left(\frac{1}{2r}\right)}
^2\right]
$$
$$
\Phi_{3}(r)=-1+{\left(1-\frac{1}{r}\right)}^{\frac{1}{2}}
~;~\Phi_{4}(r)=\frac{1}{2}ln{\left(1-\frac{1}{r}\right)}
\eqno{(1)}
$$ 
where $r$ is the radial co-ordinate scaled in units of
Schwarzschild radius. Hereafter,
 we will define the Schwarzschild radius $r_g$ as
$$
r_g=\frac{2G{M_{BH}}}{c^2}
$$
(where  $M_{BH}$  is the mass of the black hole, $G$
is universal gravitational
constant and $c$ is velocity of light in vacuum) so that the marginally bound
circular orbit $r_b$ and the last stable circular orbit $r_s$
take the values $2r_g$
and $3r_g$ respectively for a typical Schwarzschild black hole. Also,
total
mechanical energy per unit mass on $r_s$ (sometimes called
`efficiency' $e$) may be computed as $-0.057$ for this
case. We will use a simplified geometric unit throughout this paper where
radial distance $r$ is scaled in units of $r_g$, radial dynamical
velocity
$u$ and polytropic sound speed $a$ of
the flow is scaled in units of $c$ (the
velocity
of light in vacuum), mass $m$ is scaled in units of $M_{BH}$
and all other derived quantities would be scaled
accordingly. For simplicity, we will use $G=c=1$.
\noindent
Among the above potentials, $\Phi_1(r)$ was introduced by 
Paczy\'nski and Wiita (1980) which accurately reproduces the positions 
of $r_s$ and $r_b$. Also the Keplerian distribution of angular
momentum obtained using this potential is exactly the 
same as
that obtained in a pure
Schwarzschild geometry.
$\Phi_2(r)$ was proposed by 
Nowak and Wagoner (1991) to approximate
some of the
dominant relativistic effects of the accreting
black hole (slowly rotating or
non rotating) via a modified Newtonian potential. It has the correct form of $r_s$,
and it produces the best approximation for the
value of the
angular velocity $\Omega_s$
(as measured at infinity) at $r_s$ and the
radial epicyclic frequency $\kappa$ (for $r>r_s$).
$\Phi_3(r)$ and $\Phi_4(r)$ were proposed by 
Artemova, Bj\"{o}rnsson \& Novikov 1996, ABN hereafter,
to produce 
exactly the
same value of the free-fall
acceleration of a test particle at a given value of $r$ as is obtained
for a test particle at rest with respect to the Schwarzschild reference
frame ($\Phi_3$) and to produce 
the value of the free fall acceleration that is equal
to the value of the covariant component of the three dimensional free-fall
acceleration vector of a test particle that is at rest in the Schwarzschild
reference frame ($\Phi_4$) respectively. Hereafter, we will denote any $i$th
potential as $\Phi_i(r)$ where $\left\{i=1,2,3,4\right\}$ corresponds to
$\left\{\Phi_1(r), \Phi_2(r), \Phi_3(r),\Phi_4(r)\right\}$ respectively.\\

At this point, one may ask why we are interested in studying shock formation (and the
overall structure of the isothermal black hole accretion disc in general) for various
pseudo Schwarzschild potentials when YK have already performed shock study using complete
general relativistic frame work. The answer is mainly two-fold. Firstly, we have used, 
perhaps, a more realistic flow geometry (1.5 dimensional vertically integrated disc model) than that of YK
and we have provided a more detail dependence of various shock related quantities on
fundamental accretion parameters. Secondly, and more seriously,  
even if someone can provide a completely satisfactory
model for shock formation in full general relativistic 
isothermal black hole accretion, still the importance of our
work may not be
under estimated as we believe. Rigorous investigation of some  of the
shock related phenomena is extremely difficult (if not completely
impossible) to study using full general relativistic framework. Hence one
is expected to always rely on these pseudo-potentials because of the
ease of handling them. 
Also, there are possibilities that in future someone
may come up with a pseudo-Schwarzschild potential better than
any of the potentials used in this paper, 
which will be the best approximation for complete general relativistic
investigation of multi-transonic shocked flow. In such case, if one
already formulates a generalized model for multi-transonic shocked isothermal
accretion disc
for any arbitrary black hole potential, exactly what we have done in this paper,
then that generalized model will be able to readily accommodate that new
$\Phi(r)$ without having any significant change in the fundamental
structure of the formulation and solution scheme of the model
and we need not
have to worry about providing any new scheme exclusively valid only for
that new potential, if any.
\section{Governing Equations}
\noindent
Hereafter we all use the notation 
$\left[{\cal P}_{2}\right]_{i}$
for a set of values of $\left\{{\lambda},T\right\}$ for 
any particular $\Phi_{i}$.
It is to be mentioned here that for spherically symmetric isothermal accretion, one can
solve the whole problem if only $T$ alone is known. A one parameter solution scheme for
Bondi (1952) type black hole accretion in all $\Phi_{i}$s has also been 
formulated quite recently
(Sarkar \& Das 2002).
For isothermal equation of state,
$$
P=\frac{{\rho}RT}{\mu}=c_s^2{\rho}
\eqno{(2a)}
$$
where $P,{\rho},T$ and $C_{s}$
are the pressure, density, constant temperature and the constant isothermal acoustic
velocity of the flow respectively, $R$ and $\mu$ being universal gas constant and the
mean molecular weight, $C_{s}$ and $T$ can be related as 
$$
C_{s}={\Theta}T^{\frac{1}{2}}
\eqno{(2b)}
$$
where $\Theta=\sqrt{\frac{\kappa{\mu}}{m_{H}}}$ is a constant, ${m_{H}}{\sim}{m_{P}}$
being the mass of the hydrogen atom and $\kappa$ is Boltzman's constant.
Following standard literature, we consider a thin,
rotating, axisymmetric, inviscid steady flow in hydrostatic
equilibrium in transverse direction. The assumption of hydrostatic
equilibrium is justified for a thin flow because for such flows, the infall
time scale is expected to exceed the local sound crossing time
scale in the direction transverse to the flow. The flow is also assumed to
posses considerably large radial velocity which makes the flow `advective'.
We also assume that the disc to be non-self-gravitating, i.e. the increase of 
$M_{BH}$ due
to the accretion process itself is negligible. 
The local half-thickness $h_{i}(r)$  of the disc for any $\Phi_{i}(r)$ can be obtained by
balancing the gravitational force by pressure gradient and can be expressed as 
$$
h_{i}(r)=\Theta{\sqrt{\frac{rT}{\Phi_{i}^{\prime}{(r)}}}}
\eqno{(3)}
$$
where $\Phi_{i}^{\prime}{(r)}=\frac{d\Phi_{i}{(r)}}{dr}$.
The radial momentum conservation equation and the equation of 
continuity could be written as:
$$
u\frac{{d{u}}}{{d{r}}}+\frac{1}{\rho}
\frac{{d}P}{{d}r}+\frac{d}{dr}\left[\Phi^{eff}_{i}(r)\right]=0
\eqno{(4a)}
$$
$$
\frac{d}{dr}\left[u{\rho}rh_i(r)\right]=0
\eqno{(4b)}
$$
where $\Phi_i^{eff}(r)$ is the effective potential which can be defined as the summation
of the gravitational potential and the centrifugal potential for matter accreting under
the influence of $\Phi_i$ and can be expressed as:
$$
\Phi_i^{eff}(r)=\Phi_i(r)+\frac{\lambda^2(r)}{2r^2}
\eqno{(5a)}
$$
However, for inviscid flow, $\lambda(r)$ has been assumed to be $\lambda$ only.
One can calculate the expression for the effective potential in full general relativity
(D02):
$$
\Phi^{eff}_{GR}(r)=r\sqrt{\frac{r-1}{r^3-{\lambda}^2\left(1+r\right)}}-1,
\eqno{(5b)}
$$
to show that unlike $\Phi_i^{eff}(r)$, $\Phi_{GR}^{eff}(r)$ cannot be obtained by linearly
combining the gravitational and rotational energy contributions because various energies in
general relativity are combined together to produce non-linearly coupled new terms. Using
eq. (2a - 2b) and (3), we integrate eq.(4a-4b) to obtain two conservation equations for our
flow as:
$$
\frac{u_e^2(r)}{2}+{\Theta}Tln{\rho_e(r)}+\frac{\lambda^2}{2r^2}+\Phi_i(r)={\bf C}
\eqno{(6a)}
$$
$$
{\dot M}_{in}={\Theta}{\rho_e}(r)u_e(r)r^{\frac{3}{2}}\sqrt{\frac{T}{\Phi_i^{\prime}(r)}}
\eqno{(6b)}
$$
Where {\bf C} is a constant and ${\dot M}_{in}$ is the mass
accretion rate. The subscript $`e'$ indicates the values
measured on the equatorial plane of the disc; however,
we will drop $`e'$ hereafter if no confusion arises in
doing so and will denote ${\dot M}_{in}$ by ${\dot M}$. \\

For a particular value of
$\left[{\cal P}_{2}\right]_{i}$, it is quite straight
forward to derive the space gradient of the dynamical
flow velocity for any $\Phi_i$ as:
$$
\frac{du}{dr} = \left[ \frac{\left(\frac{3\Theta^{2}T}{2r}+\frac{\lambda^{2}}{r^{3}} \right) 
- \left(\frac{1}{2}\Theta^{2}T\frac{\Phi_{i}^{\prime\prime}(r)}{\Phi_{i}^{\prime}(r)} 
+ \Phi_{i}^{\prime}\right)}{\left(u-\frac{\Theta^{2}T}{u} \right)}\right]
\eqno{(7)}
$$
where $\Phi_{i}^{\prime\prime}=\frac{d^2\Phi_i}{dr^2}$.
Since the flow is assumed to be smooth everywhere, if
the denominator of eq. (7)  vanishes at any radial distance
$r$, the numerator must also vanish there to maintain the
continuity of the flow. One therefore arrives at the so
called `sonic point (alternately, the `critical point')
conditions'  by simultaneously making
the numerator and denominator of eq. (7) equal zero.
The sonic point conditions can be expressed as:
$$
u_s^i=\Theta{T^{\frac{1}{2}}}=
\sqrt{
\frac{
{\Phi_i^{\prime}}{\Bigg{\vert}}_s-\frac{\lambda^2}{r_s^3}}
{\frac{3}{2r_s}-\frac{1}{2}\left(\frac{{\Phi_i}^{\prime\prime}}{\Phi_{i}^{\prime}}\right)_s}}
\eqno{(8)}
$$
where the subscript $`s'$ indicates the quantities are to be measured at the sonic
point(s). For a fixed $\left[{\cal P}_{2}\right]_{i}$, one can solve the following
polynomial of $r_s$ to obtain the sonic point(s) of the flow:
$$
\Phi_{i}^{\prime\prime}{\Bigg{\vert}}_{s} + \frac{2}{\Theta^{2}T}\left(\Phi_{i}^{\prime}
\right)_{s}^{2} - \left[\frac{3}{r_{s}\Theta} + 
\frac{2{\lambda}^{2}}{T{\Theta}^{2}r_{s}^{3}}\right]\Phi_{i}^{\prime}{\Bigg{\vert}}_{s} = 0
\eqno{(9)}
$$
Solution of above equation is analogous to solving the following 
differential  equation at
critical points $x=x_c$: 
$$
\alpha {\ddot{y}}{\Bigg{\vert}}_{x=x_{c}} + \left[ \{ \beta\psi_{1}(x) + \gamma\psi_{2}(x) \}_{x=x_{c}}
+{\dot{y}}\left|_{x=x_{c}} \right.  \right] {\dot{y}}  {\Bigg{\vert}}_{x=x_{c}} = 0
\eqno{(10)}
$$
where :
$$
y = f(x),~ {\dot{y}} = \frac{dy(x)}{dx},~ {\ddot{y}} = \frac{d^{2}y(x)}{dx^{2}},
$$
$f(x), \psi_{1}(x)$ and $\psi_{2}(x)$
are different functions of x and $\left\{{\alpha},{\beta},{\gamma}\right\}$
are either constant or various functions of initial parameters.\\
The critical value of the dynamical velocity gradient at sonic point(s) can be
computed by solving the following equation for $\left(\frac{du}{dr}\right)_{s,i}$:
$$
\left(\frac{du}{dr}\right)_{s,i} = \pm\frac{1}{\sqrt{2}} \left[ \frac{1}{2}\Theta^
{2}T {\Bigg{[}} \left(\frac{\Phi_{i}^{\prime\prime}}{\Phi_{i}^{\prime}} \right)_{s}^{2} 
- \left(\frac{\Phi_{i}^{\prime\prime\prime}}{\Phi_{i}^{\prime}} \right)_{s} \right] 
$$
$$
- \left(\Phi_{i}^{\prime\prime}\Big{\vert}_s
+\frac{3\Theta^{2}T}{2r_{s}^{2}}+\frac{3\lambda^{2}}{r_{s}^{4}} \right)
 {\Bigg{]}}^\frac{1}{2}
\eqno{(11)}
$$
\section{Multi-transonic Accretion and Wind}
\noindent
For all $\Phi_{i}$, we find a significant region of parameter space spanned by
$\left[{\cal P}_{2}\right]_{i}$ which allows the multiplicity of sonic points for
accretion as well as for wind where two real physical inner and outer (with
respect to the black hole location) $X$ type sonic points $r_{in}$ and $r_{out}$
encompass one $`O'$ type unphysical middle sonic point $r_{mid}$ in between. For a
fixed set of $\left\{{\lambda},T\right\}$, one can solve eq. (9) to obtain
required sonic points and for a particular set of values of
$\left\{{\lambda},T\right\}$, one obtains at most four real roots, one of which
is always less than one $r_g$, hence falls inside the event horizon. Thus
maximum number of real physical roots is three, hence maximum number of sonic
points for any $\left[{\cal P}_{2}\right]_{i}$ is also three. Apparently there
remains an ambiguity in determining the values of $\left[{\cal P}_{2}\right]_{i}$
for which three sonic points form in accretion and the values of 
$\left[{\cal P}_{2}\right]_i$ for which three sonic points form in wind  because just by studying the values
of $\left\{{\lambda},T\right\}$ or $\left\{r_{in},r_{mid},r_{out}\right\}$,
one cannot determine whether it is accretion or wind which is multi-transonic
for that $\left[{\cal P}_{2}\right]_{i}$. However, we can unambiguously classify
the parameter space in the following way:\\
Let us suppose that for a particular value of $\left[{\cal P}_{2}\right]_{i}$,
accretion can be multi-transonic. So the subsonic far field solution will
become supersonic after crossing the outer sonic point ${r_{out}^{acc}}$ in accretion.
It would manifest the tendency for passing through the
inner sonic point $r_{in}^{acc}$
again after a shock transition if the solution passing through
the inner sonic point is more preferred by nature for some physical reason(s).
One can show that it is the total energy content of the flow which will
discriminate the solutions passing through ${r_{out}^{acc}}$ and
${r_{in}^{acc}}$. If the energy content of the solution passing through
$r_{in}^{acc}$ is lower than the solution passing through ${r_{out}^{acc}}$, then
accretion flow will prefer to pass through $r_{in}^{acc}$ because nature will
always prefer a lower energy solution. The whole situation would be reversed
for multi-transonic winds, energy content of wind solutions through $r_{in}^{wind}$
will be {\it higher}  than that through $r_{out}^{wind}$. One can calculate the energy
difference between solutions passing through $r_{out}$ and $r_{in}$ as:
$$
\delta{\epsilon} = 
\epsilon_{out}-\epsilon_{in}=
\frac{\lambda^{2}}{2} \left( \frac{1}{r_{o}^{2}} -\frac{1}{r_{i}^{2}} \right) 
+ \left(\Phi_{i}\left|_{r_{o}}-\Phi_{i}\right|_{r_{i}}\right)
$$
$$+
ln{ \left[ \left( \frac{r_{i}}{r_{o}} \right)^{\frac{3}{2}}
\left( \frac{\Phi_{i}^{\prime} \left|_{r_{o}} \right.}{\Phi_{i}^{\prime}\left|_{r_{i}}\right.} \right)^{\frac{1}{2}} \right]}^{\Theta^{2}T}
\eqno{(12)}
$$
where $r_o=r_{out}$ and $r_i=r_{in}$.
For a particular $\Phi_{i}$, if ${\cal A}_i\left[{\cal P}_2\right]_{i}$
denotes
the universal set representing the entire parameter space covering all values
of $\left[{\cal P}_{2}\right]_{i}$, and if ${\cal B}_i\left[{\cal P}_2\right]_{i}$ 
represents one particular subset of ${\cal A}_i\left[{\cal P}_2\right]_{i}$ 
which contains only the particular values of $\left[{\cal P}_2\right]_{i}$ 
for which eq. (9) gives three real physical roots (four
real roots with one unphysical root lying inside the event horizon), 
then ${\cal B}_i\left[{\cal P}_2\right]_{i}$ can further be decomposed 
into two subsets 
${\cal C}_i\left[{\cal P}_2\right]_{i}$ and ${\cal D}_i\left[{\cal P}_2\right]_{i}$ 
such that 
$$
\cpi{\subseteq}\bpi~{\rm only~for}~{\delta}{\epsilon}~>~0
$$
and:
$$
\cpi{\subseteq}\dpi~{\rm only~for}~{\delta}{\epsilon}~<~0
$$
For ${\pii}{\in}{\cpi}$,
we get multi-transonic {\it accretion} and for ${\pii}{\in}\dpi$ one obtains
multi-transonic {\it wind}. In Fig. 1, we plot:
$$
\left(\lambda_i,T_i\right){\in}{\pii}{\in}{\cpi}{\subseteq}{\bpi}
$$
and in Fig. 2. we plot:
$$
\left(\lambda_i,T_i\right){\in}{\pii}{\in}{\dpi}{\subseteq}{\bpi}
$$
For  all $\Phi_i$, while the specific angular momentum is plotted along X axis, the
flow temperature $T_{10}$, scaled in the unit of 10$^{10}K$, is plotted along Y
axis. 
For Fig. 1,  $A_iB_iC_i$ represents the region ${\cal C}_i\left[{\cal P}_2\right]_i$
for a particular $\Phi_i$ and for Fig. 2, $A_iB_iC_iD_i$ represents the region 
${\cal D}_i\left[{\cal P}_2\right]_{i}$ for a particular $\Phi_i$. 
From Fig. 1 it is evident that no region of parameter space common to all $\Phi_i$
is found for which $\left[{\cal P}_2\right]_{i}\in{\cal C}_{i}
\left[{\cal P}_2\right]_i$. However, significant region of parameter space is
obtained for which $\left[{\cal P}_2\right]_{i}\in{\cal C}_{i}
\left[{\cal P}_2\right]_i$ for $\Phi_2$ and $\Phi_3$. It is interesting to note
that (see Fig. 2 of D02) same trend is observed for multi-transonic 
{\it polytropic} accretion as well. The maximum flow temperature in $T_{10}$ for which one can
obtain multi-transonic accretion may be denoted by $T^{max}_i$ for any $\Phi_i(r)$.
One observes that 
$$
{T}_3^{max}~>~{T}_4^{max}~>~{T}_1^{max}~>~{T}_2^{max}
$$
and the minimum value of angular momentum ${\lambda_i^{min}}$ for which one
obtains the multi-transonic accretion, can be arranged as:
$$
{\lambda}_3^{min}~<~{\lambda}_2^{min}~<~{\lambda}_4^{min}~<~{\lambda}_1^{min}
$$
Hence one can observe that for $\Phi_3$ and $\Phi_2$, even very low angular
momentum flow (Quasi-Bondi accretion, so to say) may produce multi-transonic
accretion.\\
Situation is not exactly similar for multi-transonic {\it wind}. From Fig. 2 one
observes that there is a common region for ${\cal D}_i\left[{\cal P}_2\right]_i$ 
for flows in $\Phi_1$, $\Phi_4$ and $\Phi_3$, and there exist a separate common zone
for ${\cal D}_i\left[{\cal P}_2\right]_i$ in between $\Phi_2$, $\Phi_4$ and $\Phi_3$.
Here also same sequential order is maintained for $T_i^{max}$:
$$
{T}_3^{max}~>~{T}_4^{max}~>~{T}_1^{max}~>~{T}_2^{max}
$$
and for ${\lambda_i^{min}}$:
$$
{\lambda}_3^{min}~<~{\lambda}_2^{min}~<~{\lambda}_4^{min}~<~{\lambda}_1^{min}
$$
However, there are significant difference for range of $\lambda$ for multi-transonic
accretion and wind. Unlike multi-transonic accretion, three sonic points in {\it wind}
form for considerably high values of $\lambda_i$; particularly for $\Phi_1$ and
$\Phi_4$, the value of $\lambda$ is very large. It is to be noted that for
{\it polytropic} accretion and wind, the situation was quite different (see Fig. 2 of
D02); unlike isothermal accretion and wind, it was observed there that 
$$
{\lambda}_i^{max}{\Bigg{\vert}}_{\rm accretion} {\sim}~{\lambda}_i^{max}
{\Bigg{\vert}}_{\rm wind}
$$
It is evident from Fig. 2 that multi-transonic wind is usually obtained
obtained
only for strongly
rotating flows, i.e., for flows with intense rotational energy content, and
quasi-Bondi flow does not produce multi-transonic wind, this trend is 
more explicit for flows in $\Phi_1$ and $\Phi_4$.
\section{Shock formation and related phenomena}
In this work we are not interested in shock formation in wind, rather we
will concentrate on shock formation in accretion only. For any $\Phi_i$, if
shock forms, then $\left[{\cal P}_2\right]_i$ responsible for shock
formation must be somewhere from the region $A_iB_iC_i$ (see Fig. 1), i.e,
for which $\left[{\cal P}_2\right]_i\in{\cal C}_i\left[{\cal P}_2\right]_i$,
though not all such $\left[{\cal P}_2\right]_i$ will allow shock transition.
A multi-transonic accretion solution containing shock waves may behave as
follows: \\

The far-field subsonic solution will become supersonic after
passing the outer sonic point $r_{out}$. As the specific energy content of
the solution passing through the inner sonic point $r_{in}$ is {\it less}
compared to the already supersonic solution through $r_{out}$, the solution
through $r_{out}$ will tend to `jump down' to it's counter solution passing
through $r_{in}$. This jump will take place via a shock transition and
certain amount of energy will be dissipated at shock, while the flow
temperature (hence the accoustic flow velocity) will remain the same. The post
shock flow will become subsonic and will again accelerate to supersonic as
it passes through $r_{in}$. The combined shocked flow field is thus subsonic at
infinity and supersonic near the black hole, a physically acceptable
solution for black hole accretion.\\
Like temperature, the baryon flux, the momentum flux and a specific function
of Mach number are also conserved at shock. Following C89, those
conservation equation could be written as:
$$
\left[{\rho}M{\Theta}T^{\frac{1}{2}}\right]=0,
\left[P+{\rho}u^2\right]=0,
\left[M+\frac{1}{M}\right]=0
\eqno{(13)}
$$
where $M$ denotes the Mach number, and $\rho$ and $P$ are the vertically
integrated values of density and pressure. Also 
$$
\left[\alpha+\beta\right]{\equiv} \left\{\left(\alpha+\beta\right)_{-}-
\left(\alpha+\beta\right)_{+}\right\}
$$
$$
\left[\alpha\beta\right]\equiv\left\{{\alpha_-}{\beta_-}-{\alpha_+}{\beta_+}\right\}
$$
as described in \S 1.
Using eq. (13), we can formulate a generalized shock condition, which is the
following
$$
\frac{1}{{\rho}_{-}}-\frac{u_-^2+{\Theta}^2T}
{u_-^2{\rho}_-+\left(2{\rho}_+-{\rho}_-\right){\Theta}^2T}=0
\eqno{(14)}
$$
The above equation is valid {\it only} at the shock location, ``+" and ``-"
refers to the post and pre shock quantities respectively. For the kind of
shock discussed here, the shock strength ${\cal S}_{i}$ (which is basically the
ratio of the pre- to post-shock flow Mach number) is exactly equal to the
shock compression ratio $R_{comp}$ (the ratio of vertically integrated 
post- to pre-shock flow
density).
$$
{\cal S}_i=R_{comp}
\eqno {(15)}
$$
The amount of energy dissipation at shock can be computed as 
$$
{\Delta}{\epsilon}=
\frac{{\dot M}^2{\Phi_i^{\prime}}{\Bigg{\vert}}_{r=r_{sh}}}
{{\Theta}^4T^2{\rho}_+^2r_{sh}^3}
\left(R_{comp}^2-1\right)-\Theta^2Tln\left(R_{comp}\right)
\eqno {(16)}
$$
where $\dot M$ is the accretion rate scaled in the unit of Eddington rate.
In another way eq. (16) can be written in $\dot M$ independent form as:
$$
{\Delta}\epsilon = \frac{1}{2}\left(M_{-}^{2} - M_{+}^{2} \right) - \Theta^{2}T ln{\cal S}_i
\eqno{(17)}
$$
Fig. 3 demonstrates few typical flow topologies of the integral curves of
motion for shocked flow in various $\Phi_i$ (indicated in the figure). While
the distance from the event horizon of the central black hole (scaled in the units of 
$r_g$ and plotted in logarithmic unit) is plotted along the X axis, the local
Mach number of the flow is plotted along the Y axis. One can easily obtain 
such a set of figures for any $\left[{\cal P}_{2}\right]_{i}$ which allow
shock transition. For all figures, ABC$_1$C$_2$D represents the transonic
accretion passing through the outer sonic point $r_{out}$ (marked as $B$) 
if a shock would not form. However, as ${\delta}{\epsilon}$ in between 
$r_{out}$ and $r_{in}$ is greater than zero, the flow will encounter a shock,
becomes subsonic and jumps on the branch J${\cal I}$F,  which ultimately hits the
event horizon supersonically after it passes through the inner sonic point
$r_{in}$ marked by $\cal I$ in the figure. The small circle with a
dot at its centre (``${\odot}$") indicates the location of the unphysical 
middle sonic
point $r_{mid}$. It is to be mentioned that by solving eq. (14) one obtains 
{\it two} formal shock locations $C_1$ and $C_2$ (vertical lines marked by
arrowheads through $C_1$ and $C_2$ represent the shock transitions). For
{\it every} $\Phi_i$ and for {\it any} $\left[{\cal P}_2\right]_i$ allowing
shock transition,  $r_{mid}$ {\it always} lies in between two shocks passing
through $C_1$ and $C_2$. In other words, the `outer' shock at $C_1$ is obtained
in between $r_{out}$ and $r_{mid}$ while the `inner' shock at $C_2$ is obtained
in $r_{mid}$ and $r_{in}$. 
For all $\Phi_i$s,
it is observed that the location of the inner and the outer sonic points
anti-correlate with $\lambda$ as well as with $T$. Hotter and
strongly rotating flows push the real physical sonic 
points towards the event horizon.
The corresponding values of
$\left\{r_{in},~r_{mid},~r_{out}\right\}$, the shock locations at $C_1$ and $C_2$ and the
corresponding shock strengths are indicated at the top of each figure, while
the corresponding values of $\lambda$ and $T$ (scaled in the unit $T_{10}~
\rightarrow~T/T_{10}(=10^{10}K)$) for
which the solutions are obtained, are indicated inside each figure. GBH
represents the `self wind' of the flow, which, in the course of it's motion
away from the black hole to infinity, becomes supersonic after passing through $r_{out}$ at 
$B$. The overall scheme for obtaining the above mentioned integral curves is as
follows:\\

For a particular $\left[{\cal P}_{2}\right]_{i}$, we first compute $r_{in}$,
$r_{mid}$ and $r_{out}$ by solving eq. (9). Then we obtain the dynamical
velocity gradient of the flow at sonic points by solving eq. (11). We then
calculate the local dynamical flow velocity $u(r)$, the local radial Mach number
$M(r)$, the fluid density $\rho(r)$ and any other related dynamical or
thermodynamic quantities by solving the eq. (6 - 11) from the outer sonic point
using fourth order Runge-Kutta method. We start integrating from $r_{out}$ in
two different directions. Along BH,  we only solve for $u(r)$ and $M(r)$, because
shock does not form in subsonic flows. Integration along BC$_1$C$_2$D involves a
different procedure. Along BC$_1$C$_2$D, along with our calculation of $u(r)$
and $M(r)$, we also, at every integration step (with smallest possible
step size), keep checking whether eq. (14) is being satisfied at that point.
Along the curve $\cal I$J we calculate $\rho_+$ and $u_+$. At the same time,
along the curve BC$_1$C$_2$D, from the initial point whose $r$ coordinate
is equal to the $r$ coordinate of $J$, to the final point whose $r$ coordinate
is equal to $r_{in}$, we calculate $\rho_-$. Then we substitute these values of
$\rho_-$, $\rho_+$ and $u_-$ in eq. (14) and check for which $r$ coordinate the
values of $\rho_-$,$\rho_+$ and $u_-$ satisfy the equation. We then pick up
that particular point as our shock location $r_{sh}$ and calculate the value of
${\cal S}_{i}$ and $R_{comp}$ for the shock location(s).\\

Like C89 and YK, we also obtain {\it two} real physical formal shock locations
$C_1$ and $C_2$. C89 obtained exactly same shock strength for all formal shock
locations. It is important to note that their conclusion is {\it not} true. For
{\it all} $\Phi_i$ and for all values of $\left[{\cal P}_{2}\right]_{i}$
allowing shock formation, we obtain {\it distinctively different} shock strength for
shocks through $C_1$ and $C_2$. We did not follow the stability analysis
provided by C89 because that kind of stability analysis cannot remove the
degeneracy of shock locations, it can only partially remove the multiplicity.
We follow a more robust stability analysis provided by YK where a {\it unique}
stable shock location was obtained. Following YK, we find that only the shock
through $C_1$, i.e. the `outer' shock which forms in between $r_{out}$ and
$r_{mid}$ is stable, while the other one (shock through $C_2$) is not.
It is important to note that Nakayama (1992) also
made important 
contribution to remove the degeneracy of multiple shocks in relativistic
flow.
Hereafter, we will deal with only the stable `outer' shock location.
In Fig. 4 we represents the {\it entire} parameter space spanned by
$\left[{\cal P}_{2}\right]_i$ which allows shock formation for {\it all}
$\Phi_i$. The specific flow angular momentum $\lambda$ is plotted along the X
axis while the flow temperature $T$ (scaled in the unit of
$T_{10}{\rightarrow} 10^{10}K$) is plotted along the Y axis. For any
pseudo-potential $\Phi_i$, A$_i$B$_i$C$_i$  represents the parameter space for which
$\left[{\cal P}_2\right]_i\in{\cal C}_i\left[{\cal P}_2\right]_i$. The region
bounded by $D_iB_iC_i$ and marked by S$_i$
 represents the parameter space for which the stable
standing shock forms for that particular $\Phi_i$ and the region
A$_i$B$_i$D$_i$ represents the region for which accretion is multi-transonic
yet no stable standing shock forms in accretion.\\
If $\pii{\in}{\epi}{\subseteq}{\cpi}$ represents the parameter space for 
stable standing shock formation, obviously:
$$
D_iB_iC_i{\equiv}\pii{\in}{\epi}{\subseteq}{\cpi}
\eqno{(18)}
$$
One can also define $\pii{\in}\fpi$ which is complement of $\epi$ related to
$\cpi$ so that:
$$
A_iB_iD_i{\equiv}\left[\fpi{\Bigg{\vert}}\pii{\in}\cpi,~\pii{\notin}\epi\right]
\eqno{(19)}
$$
The shock location becomes imaginary in
${\cal F}_i\left[{\cal P}_{2}\right]_i$ hence no stable shock forms in the region
A$_i$B$_i$D$_i$ even if for ${\pii}\in$A$_i$B$_i$D$_i$ produces multi-transonic
accretion. Hence one should remember that although
the multiplicity of sonic points in
accretion is a necessary condition for the existence of shocks, 
but is {\it not} a
sufficient one. \\
We calculate $r_{sh}$,${\Delta}{\epsilon}$,$S_i$,$R_{comp}$ and
all {\it possible} shock related quantities for
$\left[{\cal P}_{2}\right]_{i}\in$D$_i$B$_i$C$_i$ for all $\Phi_i$s.
Hence our calculation is much more general one compared to C89 or YK. 
In Fig. 5, we show the variation of $r_{sh}$
(plotted along the Z axis) with
$\lambda$ (plotted along the X axis) and $T$
(scaled in the unit of $T_{10}$ and plotted along the Y axis) for all four
${\Phi_i}\left({\Phi_1}{\rightarrow}(a),
{\Phi_2}{\rightarrow}(b),
{\Phi_3}{\rightarrow}(c),
{\Phi_4}{\rightarrow}(d)\right)$.\\
One observes that the maximum possible value of shock location is obtained for
flows in $\Phi_1$ while $\Phi_3$ allows minimum span in the value of $r_{sh}$.
In Fig. (6), we plot the variation of the energy dissipation at shock, i.e, the
value of ${\Delta}{\epsilon}$ (plotted along Z axis) with the flow angular
momentum $\lambda$
(plotted along the X axis) and the flow temperature $T$ (scaled in the
unit of $T_{10}$ and plotted along the Y axis) for all four pseudo potentials
marked in the figure as $\Phi_1{\longrightarrow}(A),
\Phi_2{\longrightarrow}(B),\Phi_3{\longrightarrow}(C), \Phi_4{\longrightarrow}(D)$.
The shock location correlates with $\lambda$ as well as with $T$.
This is obvious because a rapidly
rotating flow will generate a 
centrifugal barrier at a relatively
large distance
from the event horizon and a hotter flow will generate sufficient 
amount of thermal pressure to break the flow well away from the black hole.
Also it is observed that the stronger the shock (shock with large
compression ratio), the higher the amount of
dissipated energy at the shock location and that a 
shock formed closer to the
event horizon will result in a 
larger amount of energy dissipation. 
This is also obvious because the closer the shock location to the
event horizon, the higher the amount of available gravitational 
potential energy to be dissipated at the shock.

\section{Concluding Remarks}
\noindent
The
existing literature on shock formation in black hole accretion discs are heavily inclined
towards the
study of polytropic accretion and non-dissipative shocks while the formation of
dissipative shocks in an isothermal flow is a relatively less well studied subject.
Such a `preferential investigation' is perhaps due to the fact that global
isothermality in `black hole accretion' is difficult to achieve for realistic
flows (it is difficult to maintain exactly the 
same temperature throughout a disc),
the more realistic kind of accretion is expected to be described using a 
polytropic
equation of state. However, one should never forget that the techniques developed
to study disc geometry and shock dynamics in all existing literature
(including the present paper) are heavily based on various assumptions, and hence,
are far from being sufficiently
complete and self-consistent to model the exact physical
situation. Even if the vertically integrated 1.5 dimensional disc model developed
by Matsumoto et. al. (1984), which has been used in this paper, is believed to be
one of the most successful disc models, it suffers from a number of limitations in
studying the polytropic shock dynamics. One of such major limitations is, even
for an extremely thin shock, i.e. a polytropic shock with effective
zero thickness, the
thickness of the flow (i.e. the disc thickness) changes abruptly at the shock
(pre and post shock half thickness of the polytropic flow is quite different) and
it is impossible to write down the exact pressure balance condition at the shock
by vertical averaging. On the other hand,
for isothermal flows with
dissipative shocks (the kind of shock dynamics studied in this paper) at constant
acoustic velocity, the flow thickness remains the same (because of the quick
removal of dissipated energy) and a correct pressure balance condition is
obtained. Hence we believe that the study of shock formation in isothermal flow
is quite important in its own right and we have provided {\it for 
the first time} a generalized
formalism to study such shocks for {\it all}  available pseudo-Schwarzschild black hole
potentials.
Also we have studied
the {\it complete}  disc dynamics for multi-transonic accretion and wind and have
demonstrated {\it all}  the shock solutions and successfully studied the dependence of
important shock quantities on every fundamental accretion parameter. However, in this
work is has not been possible to compare the different potentials in connection to
the suitability of mimicking the exact general relativistic solutions as was 
done in D02. It was beyond the scope of this paper to formulate and solve
the complete general relativistic set of equations governing the isothermal
accretion and wind, we hope to present that work somewhere else.\\

It is to be mentioned here that a number of recent works study the
dynamical and radiative properties of the post-shock region of advective
accretion discs (inner part of the disc, so to say) in order to understand the
spectral properties of the black hole candidate (Shrader \& Titarchuk 1998, and
references therein) and
to theoretically explain a number of diverse phenomena which are supposed to take
place in the inner part of the shocked
black hole accretion discs. (Titarchuk, Lapidus \&
Muslimov 1998, and references therein, Das 1998, Das \& Chakrabarti, 1999, Das
2003, Das, Rao \& Vadawale, 2003, Naik et.al. 2001). The importance of studying
the physics of post-shock accretion flow has also been realized from
observational evidence (Rutledge et. al. 1999, Muno, Morgan \& Remillard 1999, Webb
\& Malkan 2000, Smith, Heindl and Swank 2002). Thus we believe that our
generalized model presented in this paper may have far reaching consequences in
handling various shock related phenomena because this model deals with all
available black hole potentials.
Nevertheless, one limitation of our model is that we have assumed the flow to be
totally inviscid and viscous dissipation of azimuthal angular momentum was not
taken into account. Exact modeling of viscous multi-transonic black hole
accretion, including proper heating and cooling mechanisms, is quite an arduous
task, and  we did not dare to attempt that
in this paper. However, qualitative
calculations show that the introduction of viscosity via a radius-dependent 
power-law distribution for angular momentum only weakens the strength of the centrifugal
barrier and pushes the shock location closer to the event horizon, keeping 
the overall basic physics concerning the shock dynamics and related 
issues unaltered; details of
this work is in progress and will be discussed elsewhere.\\

In our work, some of the examples of the shocked multi-transonic flow, especially
for flows in $\Phi_{2}$ and $\Phi_{3}$, contain low intrinsic angular momentum.
Such weakly rotating flows may be found in nature in systems like detached or
semi-detached binaries fed by accretion from OB stellar winds (Illarianov \&
Sunyaev 1975, Liang \& Nolan 1984), semi-detached low-mass non-magnetic binaries
(Bisikalo et al. 1998) or super-massive black holes fed by accretion from slowly
rotating central stellar clusters (Illarionov 1988, Ho 1999 and references
therein). \\

It is now a well established fact that a number of galactic black hole candidate
show $QPO$  behaviours. Associated large amplitudes and short time scales of such
intensity variation suggests that such
oscillations are a diagnostic of the accretion processes in the
innermost regions of the accretion discs of the central accretor (van derkliss,
1989, and references therein)
and it has been suggested in recent years that such $QPO$s may be
generated because of the shock oscillation in black hole accretion discs
(Molteni, Sponholz \& Chakrabarti 1996,
Titarchuk, Lapidus  \& Muslimov 1999)
We apply our shock solutions obtained in this paper to compute the frequency
of the QPOs generated by such shock oscillations (Das 2003) and study the
dependence of such frequencies on various fundamental accretion and
shock parameters. \\

We would like to complete our discussion by pointing out one important finding in 
our work.
Since the accretion flows discussed in this paper are chosen
to be inviscid and non-dissipative, most of the binding energy of the flow is
dissipated at the shocks. 
Thus such shocks can serve as the efficient provider of effective
radiative cooling mechanism in the accretion discs.
From Fig. 6 it is evident that for some region of
parameter space spanned by $\left[{\cal P}_{2}\right]_i$, the amount of 
energy dissipated at the shock could be as high as more than ten percent  of
the total rest mass energy of the flow. This large amount of energy has to
be immediately removed from the accretion disc (at least from the equatorial
plane) to maintain the isothermality of the flow. This dissipated energy may
be drained from the disc through various processes
like a sudden X-ray burst (flare) of energy from the accretion discs
close to the event horizon (few $r_g$ away)
or through 
energetic leptonic/ baryonic outflows and jets, the
formation of which 
is likely to be initiated by this huge amount of dissipated energy at the 
shock.
Observational signature of such flares from the close vicinity of our
Galactic centre black hole has already been reported quite recently
(Baganoff et al. 2001). The observed length scale of the accretion flow
from where such flares are generated, are in fairly close agreements 
with our theoretical calculations presented in this paper.
Thus one 
concludes that the kind of shock discussed in this paper could be efficiently
radiative and `bright'. The energy dissipation at the shock and it's channeling
from the accretion disc system may explain some interesting related issues like
shock generated X-ray flares and the
QPO of galactic microquasars (Das \& Rao 2003,
in preparation) and the
generation mechanism of accretion powered galactic jet (Das 2003a,
in preparation).

\acknowledgments
\noindent
This research has made use of NASA's Astrophysics Data System Bibliographic Services. 
Research of TKD at UCLA 
is supported by NSF funded post doctoral fellowship
(Grant No. NSF AST-0098670). SM acknowledges the 
financial assistance from CSIR (India) in the form of graduate student fellowship. JKP
would like to acknowledge the kind hospitality provided by IUCAA through the 
visiting student research program.
Authors owe much debt to Prof. Mark R. Morris for reading the 
manuscript extremely carefully and for providing a number of
useful suggestions.
Authors would also like to thank the anonymous referee and the editor Prof. 
Luigi Stella for useful comments, and Atashi Chatterjee 
for typing the manuscript.

{}
\newpage
\begin{center}
{\large\bf Figure Captions} \\[1cm]
\end{center}
\noindent
Fig.\ 1: Parameter space division for multi-transonic accretion in four different
pseudo-Schwrzschild potentials marked in the figure. Specific angular momentum of
the flow ($\lambda$) is plotted along the X axis while the flow temperature is
plotted along the Y axis in the unit of 10$^{10}K$. Note that there is no common
set of $\left\{\lambda,T\right\}$ for which multi-transonic flow occurs for all
potentials, see text for details.

\noindent
Fig.\  2: Multi-transonic isothermal wind as a function of 
$\left\{\lambda,T{\longrightarrow}T/T_{10}\left(=10^{10}K\right)\right\}$
for various black hole potentials marked on the figure. Note how, unlike
accretion, multi-transonicity is obtained in wind for very strongly rotating
flows (flow with large value of $\lambda$). 

\noindent
Fig.\ 3: Solution topology of multi-transonic shocked accretion along with it's 
`self-wind' for four different black hole potentials marked in the figure. Among
the two formal shock solution (marked by vertical lines with downward arrow
having {\it different} values of shock strength, solution through $C_1$ comes out to be
the stable one, see text for detail. 

\noindent
Fig.\  4: Global parameter space for accretion solution with shocks (region marked
by $S_1$ and bounded by $B_1C_1D_1$ for the potential $\Phi_1$ and so on) embedded
in region for general multi-transonic accretion solutions (region bounded by 
$A_1B_1C_1$ for $\Phi_1$ and so on) for four different potentials. This figure
demonstrates {\it all} possible
$\left\{\lambda,T{\longrightarrow}T/T_{10}\left(=10^{10}K\right)\right\}$
for which shock may form in isothermal black hole accretion disc. 

\noindent
Fig.\  5: Variation of shock location $r_{sh}$ (scaled in the unit of Schwarzschild
radius and plotted along the Z axis) with specific angular momentum (plotted
along the X) axis and flow temperature $T$ (scaled in the unit of
$T_{10}=10^{10}K$ and plotted along the Y axis) for four potentials 
$\Phi_{1}(a),\Phi_{2}(b),\Phi{3}(c)$ and $\Phi_{4}(d)$. See text for detail. This
figure is complementary to Fig. 4 in the sense that shock locations in this
figure are obtained for {\it all} 
$\left\{\lambda,T{\longrightarrow}T/T_{10}\left(=10^{10}K\right) \right\}$ in Fig. 4
for which shock may form. 

\noindent
Fig.\ 6: Variation of the amount of energy dissipation ${\Delta}{\epsilon}$ at the
shock location (plotted along the Z axis) as a function $\lambda$ (X axis) and
$T/T_{10}\left(=10^{10}K\right)$ Y axis for four different potentials 
$\Phi_1(A),\Phi_2(B),\Phi_3(C)$ and $\Phi_4(D)$, see text for detail.   
\newpage
\plotone{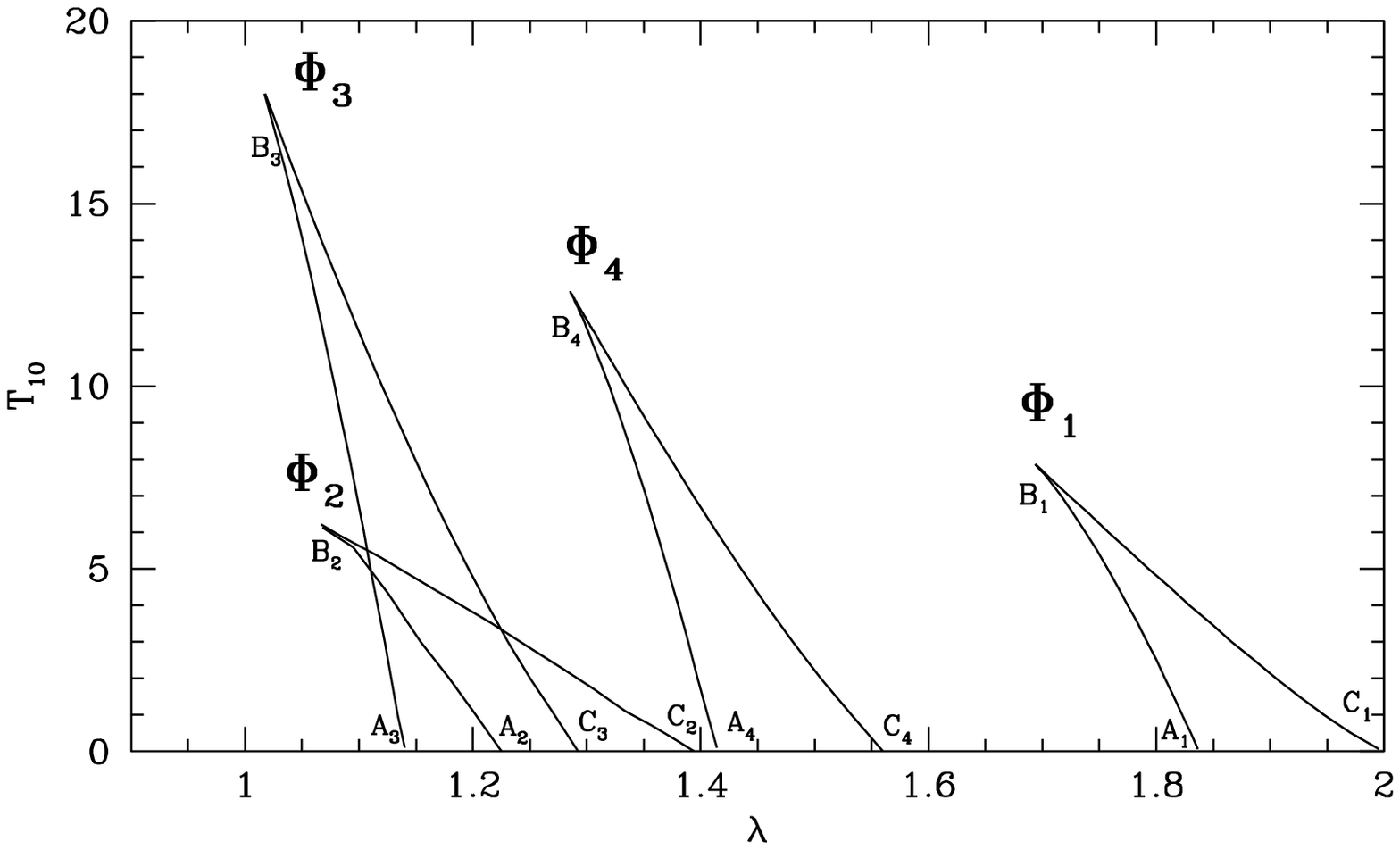}

\newpage
\plotone{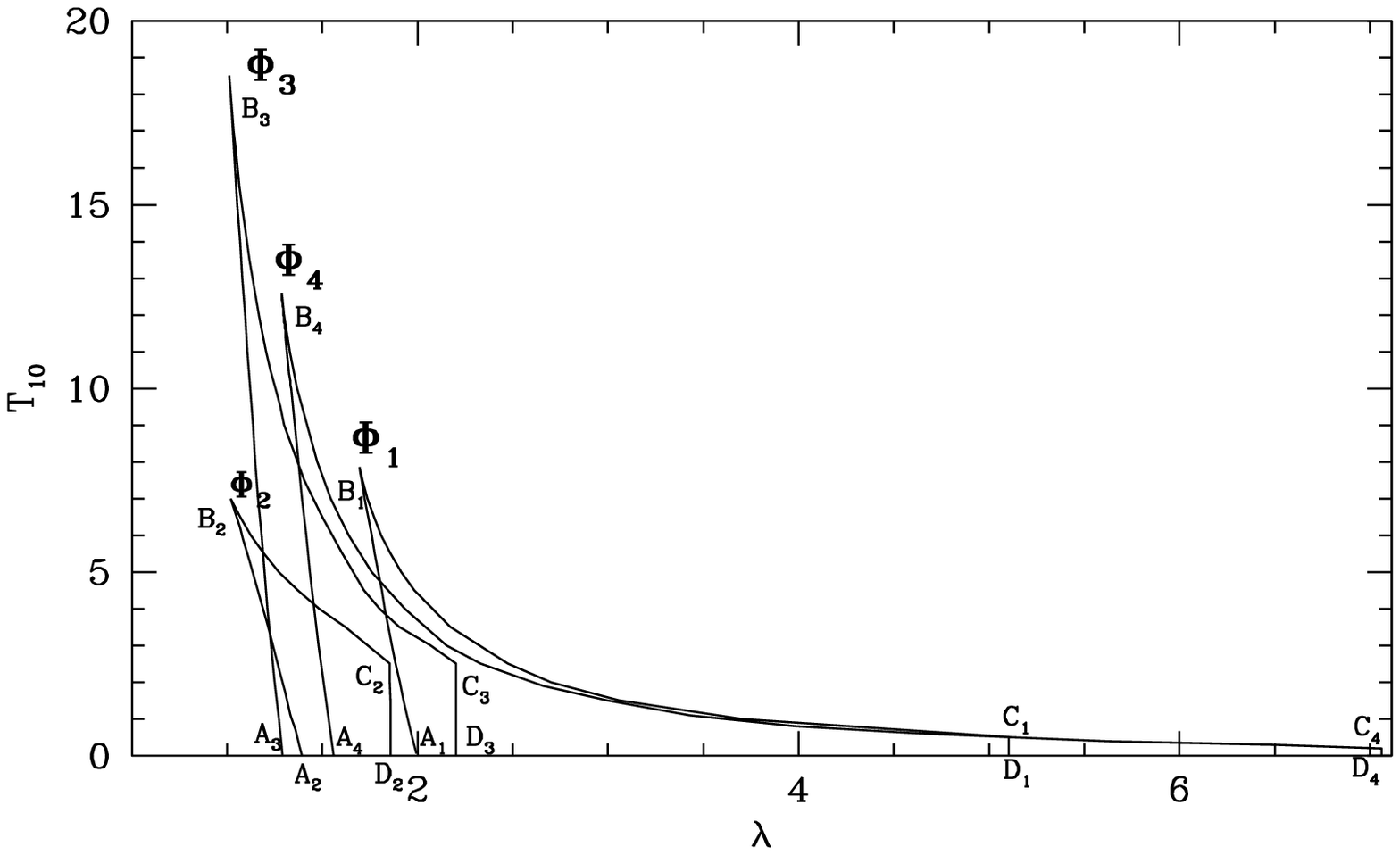}

\newpage
\plotone{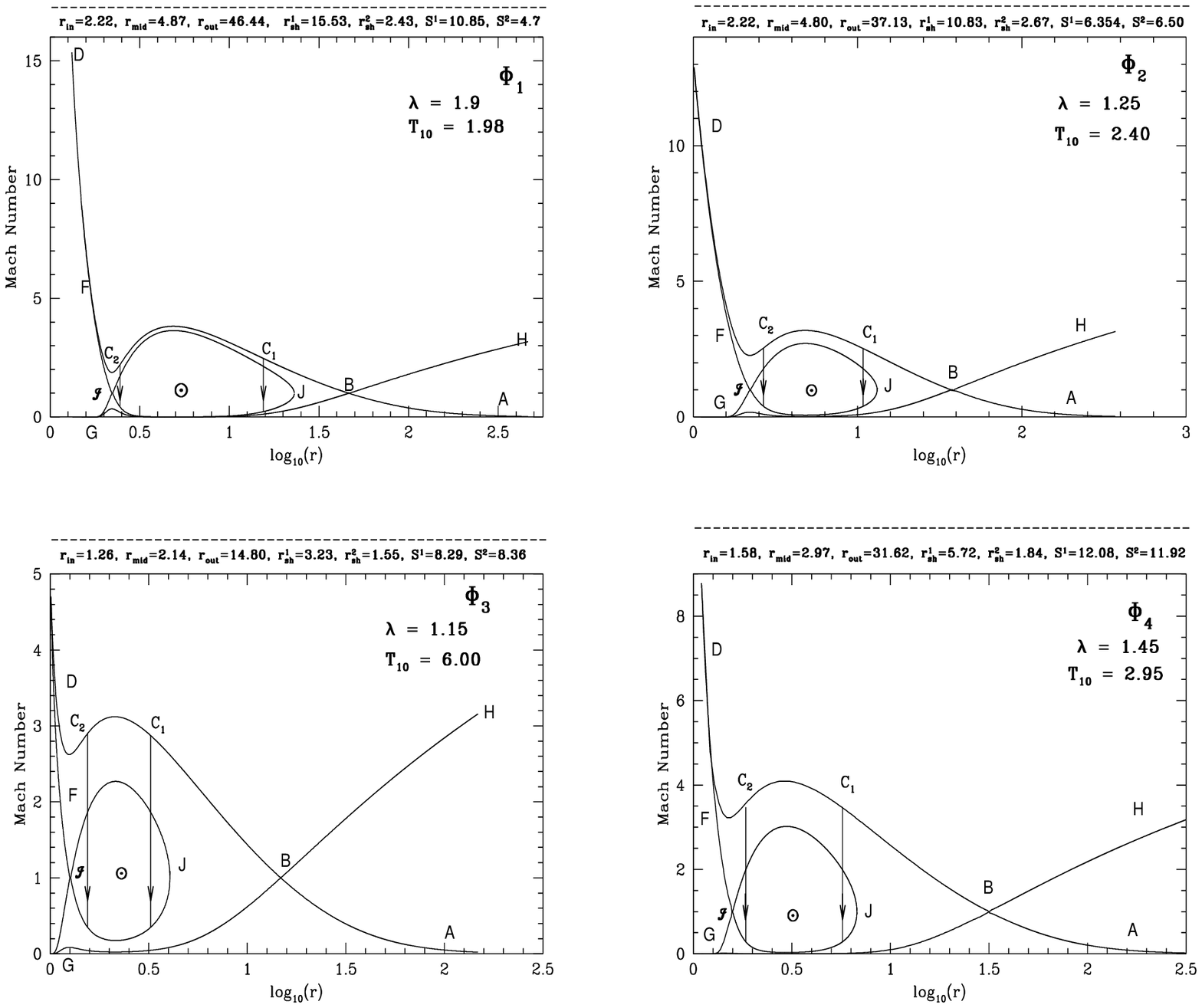}

\newpage
\plotone{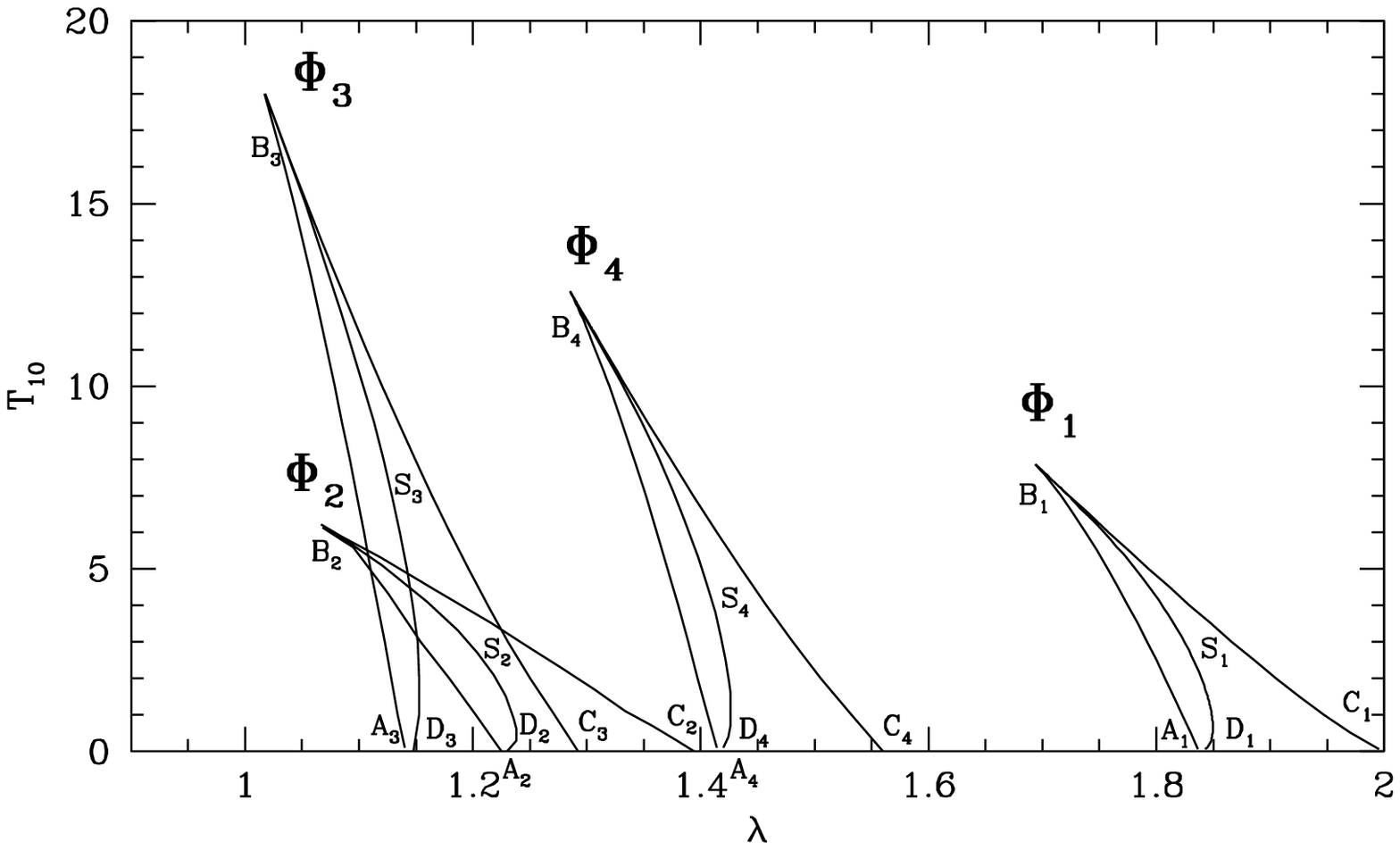}

\newpage
\plotone{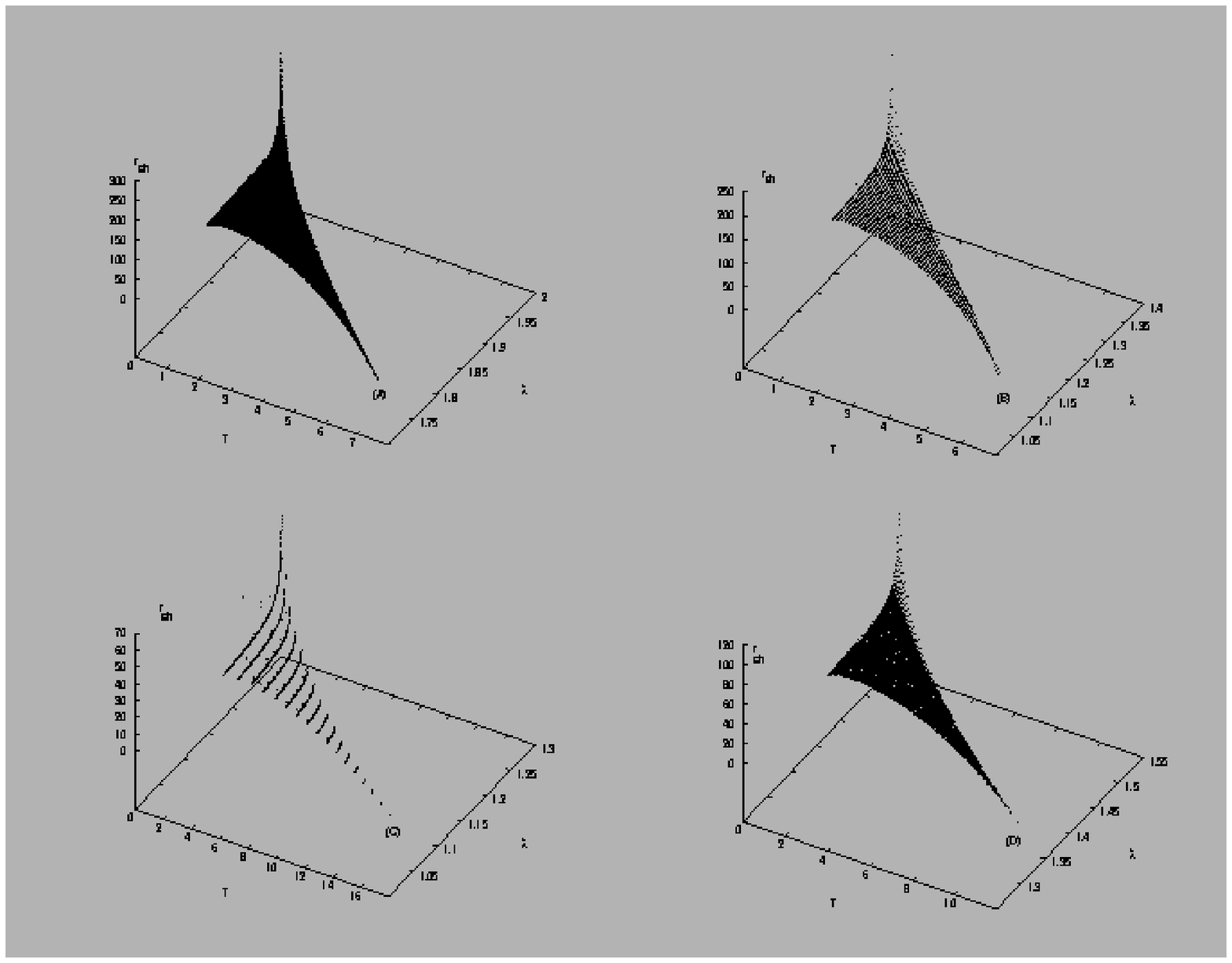}

\newpage
\plotone{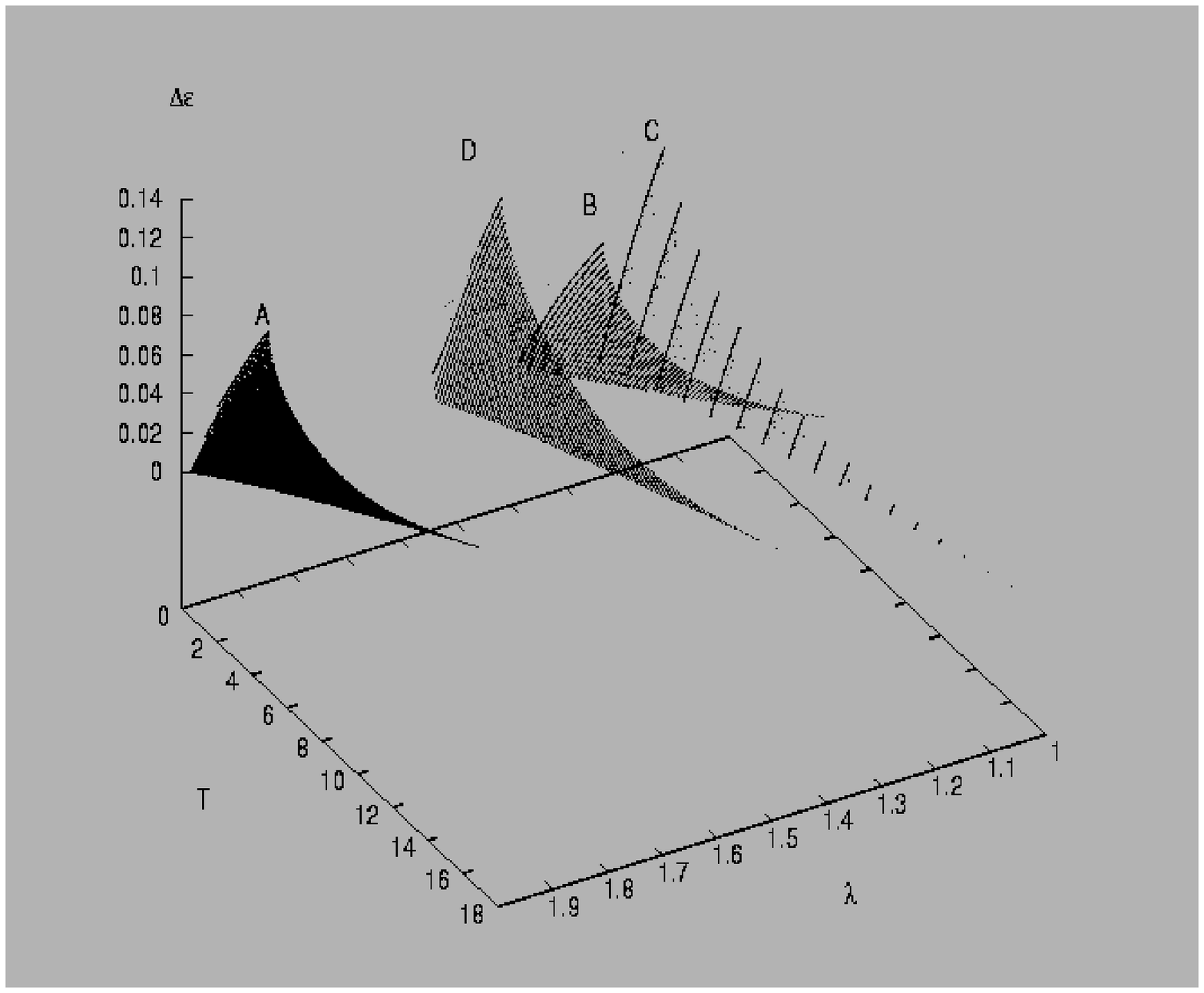}

\end{document}